\documentclass [letterpaper]{jpconf}
\usepackage{graphicx}

\begin{document}
\title{Review on possible Gravitational Anomalies}
\author{Xavier E. Amador}

\address{Centro de Investigaci\'on y Estudios Avanzados, CINVESTAV, Dept. of Physics, Av. IPN 2508, 07000 Ciudad de Mexico, D.F.,
Mexico}
\ead{xamador@fis.cinvestav.mx}

\begin{abstract} This is an updated introductory review of 2 possible gravitational anomalies that has
attracted part of the scientific community:  the Allais effect that occur during solar eclipses, and the Pioneer 10 spacecraft anomaly,
experimented also by Pioneer 11 and Ulysses spacecrafts.  It seems that, to date, no satisfactory conventional explanation exist to
these phenomena, and this {\it suggest} that possible new physics will be needed to account for them.  The main purpose of this review
is to
announce 3 other new measurements that will be carried on during the 2005 solar eclipses in Panama and
Colombia (Apr. 8) and in Portugal (Oct.15). \end{abstract}

The interest on this type of anomalous behaviours consist in that they could be a signal of slight departures from the known
gravitational
laws, which translates into new physics.  Mexico has been witness of this interest in the last 2 meetings: {\it 2nd Mexican Meeting on
Mathematical
and Experimental Physics} \cite{mex1}, D.F., Mexico, sep. 2004, where M. Nieto introduced us to the problem and in this
{\it VI
Mexican School on Gravitation and Mathematical Physics, "Approaches to Quantum Gravity"}, Playa del Carmen, Quintana Roo, Mexico,
nov. 2004.

\section{Description of the Problem}

In June 30, 1954, {\it Maurice Allais} \cite{allais} (french physicist, winner of the 1988 Nobel Prize in Economics, winner of the 1959
Galabert Prize of the French Astronautical Society, and also a laureated of the United States Gravity Research Foundation due to his
gravitational experiments)  reported that a Foucault pendulum exhibited anomalous movements at the time of a solar eclipse that cannot
be explained by
conventional effects [disturbances of an aleatory order, periodic luni-solar effects, indirect influence of some conventional factors
like seismic activity, temperature changes, atmospheric pressure, magnetism, etc.].  In October 22, 1959, he observed analogous
deviations in a {\it paraconical} pendulum that he invented.  Even the famous Werner von Braun (german
rocket scientist at NASA's Marshall Space Flight Center) encouraged Allais to publish his results.  The study of anomalous behaviours
(of pendulums) during (total and partial) solar eclipses has been repeated with torsion pendulums \cite{saxl}, and the variation in
gravity [$\sim 6 \times 10^{-8} cm/s^2$] during the eclipses has been studied with gravimeters \cite{gravim} and other devices (atomic
clocks \cite{atomic}), and some groups report positive results (supporting the existence of anomalous phenomena), but
others report no detection of the phenomena
and some has reported ambiguos results. The point is that many countries and space agencies around the world has been extremely curious
about this phenomena and has carried out several experiments to try to confirm if these are real, and to explain them in conventional or
unconventional terms (see for example: http://science.nasa.gov/newhome/headlines/ast06aug99$\_$1.htm,
http://ams.astro.univie.ac.at/$\sim$nendwich/Science/SoFi/allais2.html, \\
http://science.nasa.gov/newhome/headlines/ast12oct99$\_$1.htm).

If these anomalies turn out to be new gravitational physics, there can be a relation
to  the Pioneer anomaly and/or the anomalous rotation curves
of galaxies, as NASA has suggested (see the cited URL's). The {\it Pioneer 10} (launched in
1972) {\it anomaly} is a sunward small (radial) acceleration [$\sim 8(+3) \times 10^{-8} cm/s^2 \sim 0.1$ nano-$g$, where $g = 980 \, 
cm/s^2$] first detected in 1980, but reported in Sept. 24, 1998 after detailed analysis of the Doppler shifted radio data from the
spacecraft \cite{pio}:  the probe is moving as if it were subject to a {\it "new, unknown force"}  pointing towards the Sun.  The same
constant acceleration imparted by this force has been detected in 2 more spacecraft:  Pioneer 11 (launched in 1973) and Ulysses
(launched in 1990). All probes are following trajectories that cannot be explained by conventional physics:  based on General Relativity
computations of their trajectories under the gravitational pull of the {\it known} bodies in the solar system (the Sun playing the major
role), they are not in the position where they should be (the discrepancy amounts to approx. 248,500 miles ($\sim 400,000$ km), almost
the Earth-Moon distance). This anomalous acceleration was also detected in Galileo probe (unreliable data) but not in the Voyager. 
The European Space Agency (ESA) announced in the Cosmic Vision workshop at UNESCO, in sept. 2004, that it will consider
plans for a number of high precision experiments/missions that will test the Pioneer anomaly directly.  If this anomalous acceleration
can not be explained away by some mundane systematic effect, it would be a slight departure from the predictions of General
Relativity, i.e., new physics.  As to date, no conventional explanation has been found.  It was remarked (see \cite{duif} and cited
references) that the order of magnitude
of the gravity variations measured by the gravimeters during the solar eclipses, that of the spacecrafts anomaly and that of the
critical
acceletarion set by {\it MO}dified {\it N}ewtonian {\it D}ynamics (MOND, an alternative to Dark Matter)  \cite{mond} are all the same:
$10^{-8} cm/s^2$.

\section{Allais' effect}

A Foucault's pendulum [FP] hangs from a special joint that permits free rotation around the vertical and it can track the Earth's
rotation.  Before and after the 1954 and 1959 (partial) solar eclipses the plane of oscillation of a FP and of a "paraconical" pendulum
(one which is suspended via a small steel ball bearing, which is capable of rolling in all directions upon a plane
horizontal surface) experienced normal constant clockwise rotation, with an average angular velocity of 0.19 degrees per minute. But at
the begining of the eclipse the angular velocity of the oscillation plane changed abruptedly an average of 13.5 degrees per minute,
counterclockwise! (fig.1, a).  This huge sudden change in the sense and in the rotation of the oscillation plane persisted
throughout the length (2.5 hours) of the eclipse. Similar experiments with desfavorable conclusions were reported in \cite{para}.

\begin{figure}[th]
\begin{center}
\includegraphics[scale=0.85]{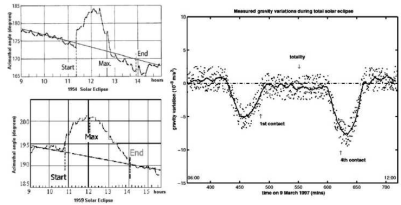}
\end{center}
\caption{a) change in the angular velocity of the oscillation plane observed by Allais; b) gravity variation measured by a gravimeter,
during a solar eclipse in 1997, see Yang and Wang in \cite{gravim}.}
\end{figure}

\subsection{Conventional possible (local) explanations}

There has been several studies \cite{duif} that try to explain this effect: seismic disturbances due to increased human activity before
and just after an
eclipse; gravitational effects of an increased density of an air mass spot (pressure changes) due to cooling of the upper atmosphere, or
due
to tidal waves on the shell of the Earth;  tilt due to temperature changes of the soil and other atmospheric influences; tilt
due to atmospheric loading; instrumental (systematic) errors during the measurement and other effects on the instruments (for example,
thermal expansion of the pendulums, etc.).$\fullsquare$

Incredible enough, all conventional mundane possibilities proposed so far do not work.  It's suspected that a {\it gravitational effect}
is the cause of this phenomenon.  Even though testing gravitational variations to verify the effects of a solar eclipse is a difficult
enterprise (due to the short duration of an eclipse it is difficult to conduct many tests), several observers in a distributed global
network of observing stations have done it, using different kinds of instruments, inside and outside buildings, far away from populated
areas, ruling out in this way instrumental errors, seismic effects, atmospheric and temperature changes, etc.  The analysis of the
pressure and temperature changes in the atmosphere indicates that these lag behind the eclipse shadow by several minutes, so it's
difficult to explain the abrupt changes observed in the instruments.  Other similar effects are too small to account for the magnitude
of the observed phenomena.  The observation of the anomalies has been repeated also using {\it torsion pendumlums} \cite{saxl}:  the
period of oscillation of the pendulum showed a considerable increase during the solar eclipse and, according to \cite{saxl}, this was
caused by a change in the tension (produced by the weight) of the wire, which means a change in the local gravitational field. There are
also reports of a gravitational variation measured by {\it gravimeters} (fig.1, b) \cite{gravim}. The gravitational variations were
about $\sim 6 \times 10^{-8} cm/s^2$ ($1 \times 10^{-8} cm/s^2 = 1$ micro-gal).  The observation with 2 {\it atomic clocks} reported
strange effects in time measurements \cite{atomic} during various solar eclipses (sept.23, 1987, mar.18, 1988, jul.22, 1990).  However,
it seems that all measurements carried out so far has not been conclusive.  Also, not all experiments report positive results.  So the
issue is still uncertain.

It was suggested \cite{duif} that a {\bf global measurement} under strictly controlled conditions is needed in order to determine
whether
these effects are real, so several issues
must be under control simultaneously during the next solar eclipse experiments/observations: a)  many observing stations
(those that lie directly under the path of the total eclipse) in different locations are needed:  it's necessary a global effort of a
worldwide network of scientists collaborating on a single eclipse, like the one (unfinished!) carried out by NASA in 1999 (see the
NASA's 
URLs) ; b)
several
and different instruments must be used simultaneously at several museums observing with Foucault pendulums, at other places using other
types of pendulums (paraconical, torsion) composed of different materials, using gravimeters, etc.; c) station's locations must be
protected/shielded against:  seismic activities, atmospheric pressure and temperature changes, and to monitor these with the appropiate
equipment, etc.

\subsection{Non Conventional explanations} If the anomalies are real, {\it why only occur during a solar eclipse?} The Sun-Moon-Earth
system is nearly aligned about once a month (on new moon).  So, {\it why nobody has noticed it before?} May be due to the smallness of
the effect and the instrumental resolution. Even though the situation is far from clear, several radical explanations have been invoked:  
a possible anisotropy of space (different properties of space in different directions), gravitational waves and solar radiation, effects
related wit MOND, Sun's gravity effects on Dark Matter flux in the Solar system \cite{DM}, Majorana {\it shielding of gravity}, effects
related to the Pioneer anomaly, etc.$\fullsquare$

If these effects are related to the spacecrafts
anomaly, then more unconventional ideas are available.  In the cited NASA's URLs, there is a list of published results by
many
scientific institutions around the world.

\section{Pioneer Anomaly}
\subsection{Possible causes}
{\bf Conventional}. Science teaches us always to look first for conventional explanations:  a) researches believed that the
anomaly is probably caused by the space probes themselves, perhaps due to heat or gas emission (outgassing or thermal radiation), or some systematic effect
associated with the spacecraft, etc.; 
b) other causes could be some subtle effects associated with the
      tracking systems, which would be of extreme importance to know
      for, if it turns out to be the cause, understanding the anomaly
      could help space navigation engineers to build better, more
      stable,  less noisy,  quieter spacecrafts with high-precision
      navigation systems for future missions; 
c) solar radiation pressure (the solar photons, by impacting the
      probes, transfer a tiny momentum to them) or  other interactions
      between the solar wind (charged particles and ions) and the
      spacecrafts;
d) possible corruption to the radio Doppler data;
e) wobbles and other changes in Earth's rotation could affect the way
      in which the signal is received; 
f) The Aerospace Corporation (http://www.aero.org/news/current/force.html)  made an independent computer based
      analysis of the motions and ruled out errors in JPL's orbital
      determination software (JPL = Jet Propulsion Lab., )  as the
      source of the anomalous, and also ruled out the following possible
      causes: unkown perturbations from gravity of the Kuiper belt and gravity from the galaxy, 
      errors in the planetary ephemeris, errors in the values of the
      Earth's orientation-precession-nutation, 
      nongravitational effects from solar pressure and control
      maneuvers,  solar wind and interplanetary medium,  nominal thermal
      radiation and plutonium half-life,  drifting clocks, unknown general
      relativity corrections and the speed of gravity effects, hardware problems at
      the tracking stations, etc. $\fullsquare$

So far, the researchers have not found that any of these effects can account for the {\it size and direction} of the anomalous
acceleration.  
Even thought in the exhaustive analysis it was considered and ruled out many possible causes, it's expected that the explanation will
involve conventional normal physics.  However, it is also considered the possiblity or insinuation of {\it 'new (overlooked) physics'},
since
{\it "no unambiguous, onboard systematic problem has been discovered"}.

{\bf Unconventional} \cite{P5} (New physics?) a) Possible modifications to Gravity (Einstein's General
      Relativity) based on modern theoretical
      ideas in Quantum Gravity suggest that it's necessary; ideas are: gravity might
      attract a little harder than expected at large distances or small
      accelerations, so dark matter is not 
      necessary or that gravity tugs slightly
      harder at things farther away, and others suggest  some  new
      things, but several models have to deal
      with the serious problem that  if the anomalous acceleration is due to
      gravity then these effects would also
      be seen in planetary motions (Earth and Mars) and the data don't
      support this, so the effect can not be universal;
b) Possible influence of  Dark Matter [DM]:   the deceleration could be
      due to the gravitational attraction of DM or that the
      sun influences the galactic flow of  cold DM  through the
      solar system \cite{DM}, although it's generally argued that these 
      explanations can't be correct because to create the measured
      anomalous acceleration, a big amount of DM is required to be in our
      solar system and it would have affected also the motions of other
      bodies and besides, DM effects are presumed to operate at galactic
      scales and there is no evidence of it influencing anything on
      smaller scales such as in our solar system;
c) Asymmetry between matter and anti-matter: gravity works
      differently on antimatter than matter;   
d) Existence of a bulk scalar field,  or the Quintessence?; 
e) Extrange new ideas:  the anomaly is not related to the motion of
      the spaceship, but it's a consequence of the acceleration of the
      cosmological proper time with respect to the coordinate parametric
      time, which is an effect of the background gravitational potential
      of the entire universe, or that the anomaly is due to the
      existence of a Mirror matter  (mirror gas or dust) in our
      solar system (and idea from the Mirror symmetry in String
      theories);
f) New ones based on attempts to go beyond the Standard Model of
      particle physics: could have a relation with  Loop Quantum Gravity
      and the inverse of the cosmological constant, which
      could establish a new distance scale, which can be used 
      to build an natural acceleration constant that has the same order
      of magnitude as the anomaly,  Strings/Branes theories in the
      context of Brane-World scenarios,  higher dimensions of
      space introduce new degrees of freedom and  could provide weak
      forces on the scale of the solar system, possible violations of space-time symmetries such
      as Lorentz symmetry; 
g) Others: consult the meeting at Center for Applied Space Technology and
      Microgravity (ZARM), University Bremen, Germany, held in May 18,
      2004 (http://www.zarm.uni-bremen.de/Pioneer). $\fullsquare$

None of the current proposal (conventional or not) can, so far, explain satisfactorily the Pioneer anomaly.  It is therefore vital to
test gravity more precisely.  Researchers at NASA's Jet Propulsion Laboratory, at the Center for Applied Space
Technology and Microgravity of the University Bremen, Germany and at the Los Alamos National Laboratory want to reanalyse earlier and
less
precise Pioneer data.  It's important to remark again that even though all this could suggest exciting new ideas and
possibilities, many of the researchers involved directly in these matters support the idea that the most likely cause of the anomalous
acceleration is a conventional one.  To date, there are numerous proposals to gather more data: it's has been proposed  that
NASA's future deep space mission to Pluto (with improved new technologies), could provide more high-quality data for investigating the
anomaly; also, the European Space Agency confirmed in the Cosmic Vision Workshop (at UNESCO,
Sept.15-16, 2004) that it's planning to launch (if funded, in 2015) a mission devoted solely for the study of this anomaly.

\section{Future developements} 
Measurements of the anomalous behaviours during solar eclipses can be done with instruments that are not so difficult to gather.  The
opportunity is now at hand at the 2005 solar eclipses in Panama, Colombia and Portugal.  Several teams are already planning  to perform
more measurements.

\ack{We want to thank Dra. Elena C\'aceres (CINVESTAV and Brown Univ.) and Dr. H\'ector Hugo Garc\'{\i}a-Compe\'an (CINVESTAV) for their
encouragement and support.}


\end{document}